\documentclass[a4paper,twocolumn]{esapub}

\usepackage{times}
\usepackage{natbib}
\usepackage{graphicx}

\newcommand{\Eq}[1]{Eq.~(\ref{#1})}
\newcommand{\Fig}[1]{Fig.~\ref{#1}}
\newcommand{\EQ}{\begin{equation}}
\newcommand{\EN}{\end{equation}}
\newcommand{\EQA}{\begin{eqnarray}}
\newcommand{\ENA}{\end{eqnarray}}

\title{Fragment Driven Magnetic Reconnection}
\author{K. Galsgaard}
\affil{NBIfAFG, Astronomical Observatory, Juliane Maries vej 31, 2100 Kbh \O, Denmark}
\author{C. Parnell}
\affil{School of Mathematical and Statistical Sciences, University of St Andrews,
St Andrews, KY16 9SS, UK}

\begin{document}


\maketitle

\begin{abstract}
The heating of the million degree, diffuse coronal plasma may be caused by a
large number of events that are too small to be identified by present days observations.
One explanation for these events could be the local interaction between magnetic flux
systems that divide space into numerous flux regions. When
such regions are independently advected by photospheric motions
the expected outcome is the formation of enhanced
current concentration at specific locations in space. Due to magnetic resistivity,
these currents dissipate and heat the plasma. In this paper, we investigate
a simple model where two, initially unconnected, flux systems are forced to interact in
response to the imposed boundary driving by solving the non-ideal 3D MHD equations
numerically. The reconnection rate of the dynamical process is determined
and compared with the corresponding rate for the potential evolution of the
magnetic field. This shows that the dynamic reconnection rate is about a
factor of two smaller than the potential (perfect, instantaneous) rate for realistic solar driving
velocities demonstrating that this three-dimensional magnetic reconnection process is fast.

The energy input for a fixed advection distance is found to be independent of the driving velocity. 
The Joule dissipation associated with the reconnection process is also found to be basically dependent on the
advection distance rather than driving velocity. This implies that the timescale for the event determines 
the effect the heating has on the temperature increase. 

Finally, the numerical experiments indicate
that the observational structure of the reconnection site changes dramatically depending
on the phase of the evolution of the passage of the two flux sources. In the initial
phase, where the sources become connected, the heating is confined to a compact region.
For the disconnecting phase the
energy gets distributed over a larger area due to the reconnected field line connectivity.
\end{abstract}

\section{Introduction}

The coronal magnetic field is a manifestation of the distribution of photospheric fluxes
and their continuous motion. By taking a magnetogram and representing each
pixel value by a point source Close et al. (2004a) 
showed that each coherent flux patch
connects to several neighbouring flux patches. This complex mapping of the
magnetic field gives rise to many {\it separatrix surfaces}; shells that divide space into different regions 
of connectivity. 
In general the distance, and also the direction, of travel of the various magnetic sources differ. The movement of any one source causes the orientation of the magnetic field at the
interface of the separatrix surfaces enclosing this flux region to change.
It is assumed that, in general, such changes lead to these surfaces attracting current \cite{Priest_ea02}.
If the current growth is significant magnetic reconnection is expected to occur releasing some of the
free magnetic energy built up by the movement of the sources. 

In the photosphere the flux sources are continuously moved around by the convective motions and stresses
like those indicated above must be taking place at many locations above the photosphere.
From detailed analysis of magnetograms, 
Hagenaar et al. (2003) showed that the recycle time of the photospheric flux is on the order of 8-19 hours. 
During this time flux sources
emerge into the photosphere, where after they undergo merging, fragmentation and annihilation resulting in 
the loss of their identity. Flux sources are also advected resulting in a complex magnetic field.
A direct consequence of all this activity is that the connectivity
of the sources must change on a timescale that is much shorter
than the turn over time of the sources themselves \cite{Close_ea04b}.
These processes all contribute locally to the heating of the coronal plasma, 
but their individual contributions may not be as
prominent and easily observed as large flares. 
In the past, numerous models have been investigated for coronal heating. Most of these concern processes where flux
emergence or annihilation releases the energy  
(e.g., Uchida \& Sakurai, 1977; Forbes \& Priest, 1984; Shibata, Nozawa \& Matsumoto, 1992; 
Dreher, Birk \& Neukirch, 1997; Yokoyama \& Shibata, 2001; Zhang \&
Low, 2002; Priest, Parnell \& Martin, 1994; Parnell,
Priest \& Titov, 1994; Birk, Dreher \& Neukirch, 1996).
In contrast, the situation where the flux is not emerged or annihilated, but instead is forced to interact with
flux from a different
source, have not been investigated in great detail. This process, which is highly likely to take place in the solar
magnetic field, was first investigated using the minimum current approach \cite{Longcope98},
where the interaction between independent flux sources in a constant background magnetic field
is investigated. This work showed that as the flux systems start to interact the free energy in the
field accumulates, in the form of a current sheet, at the separator line connecting two null points in 
the photospheric plane.  Similarly, the same is found when the sources start to disconnect again.
This scenario of independent flux systems first connecting and then disconnecting was investigated using 
numerical experiments \cite{Galsgaard_ea00}.
They showed that the initial process through which the two flux sources connect follows the same pattern,
namely that the two sources connect through separator reconnection. In the disconnection phase, however, the
reconnection occurs through a different process - separatrix-surface reconnection. This stage of the evolution 
is clearly different from the minimum current approach.

In this paper, we discuss the results of a series of numerical experiments that
are conducted to investigate these basic processes of the flux interaction in more
detail. In particular, the analysis concentrates on the reconnection process,
determining a scaling relation between the numerical reconnection rate and the comparable
rate for the potential evolution of the field structure, i.e., the perfect, instantaneous reconnection rate. 
Further, an analysis of the global energy input/output is investigated. Finally, a simple analysis of the magnetic
structure of the newly reconnected field lines has been made, visualising
the regions that would be exposed to large temperature increases as a result of the on going
magnetic reconnection.

\section{Model Setup}
\begin{figure}
\center
\includegraphics[width=0.98\linewidth]{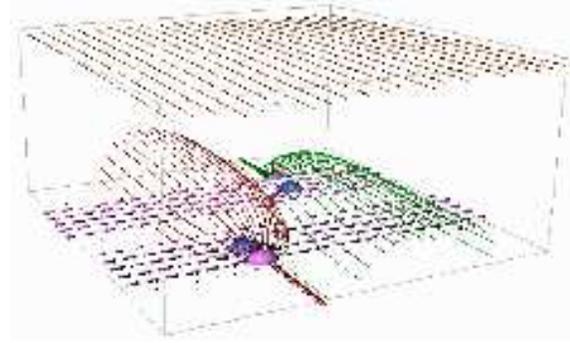}
\caption[]{\label{initial.fig} An example of the initial conditions used in the experiments. The
blue isosurfaces show the locations of the magnetic sources. The pink isosurfaces shows the location of
two magnetic null points. The brownish field lines close to the top
boundary shows the large scale magnetic field structure. The two other sets of field lines, red and green,
indicate the location of the separatrix surfaces dividing space into independent flux regions. The
arrows at the bottom boundary indicate the profile and direction of the imposed driving velocity.
}
\end{figure}

The intention with the experiment is to investigate the simplest possible situation where two
independent flux sources interact without loss (or gain) of flux from the sources. This is obtained by
setting up a situation where two, oppositely-signed, flux sources are located a short distance apart on the bottom
boundary of a Cartesian domain. To remove the dipole structure the two sources would generate if on their
own, a constant horizontal magnetic field is added. By changing the orientation of this overlying field
component the field from the independent flux domains can be changed from running parallel passed the other source, to parallel away from the other source. Introducing an advection velocity on the sources adds a preferred
direction to the setup. The advection velocity is applied such that the flux domains
of the two sources are forced to interact with each other. As they
interact strong current concentrations are generated and eventually
drive magnetic reconnection as the sources are advected passed one another. In \Fig{initial.fig}
the situation adopted in 
Galsgaard et al. (2000) is shown. Here, the isosurfaces show the locations
of the magnetic flux concentrations. The field lines starting close to the bottom boundary
indicate the separatrix surfaces dividing the space into three magnetically independent flux regions.
In comparison,
the large scale magnetic field is indicated by the field lines close to the top boundary of the
domain. On the bottom boundary, the vectors show the structure of the imposed velocity.
Notice, that the driving profile is such that it maintains the structure of the sources.

The dynamical evolution of the system is investigated by solving the non-ideal 3D MHD equations
numerically using the code described by 
Nordlund \& Galsgaard 1995.

\section{Interaction of the Flux Domains}
\begin{figure}
\center
\includegraphics[width=0.48\linewidth]{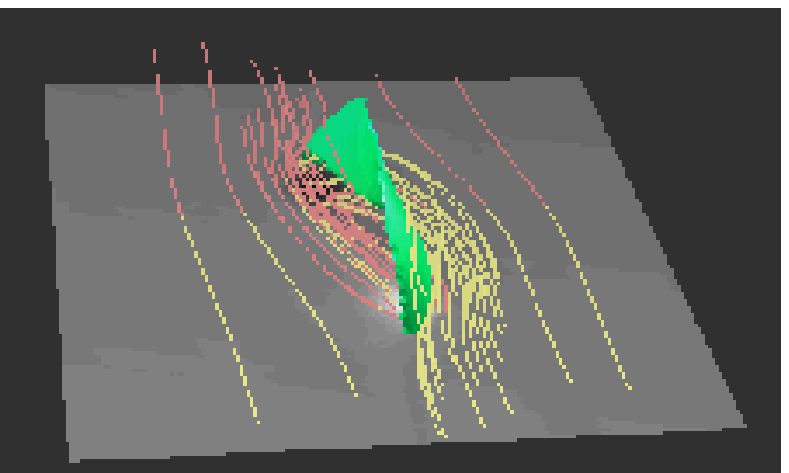}
\includegraphics[width=0.48\linewidth]{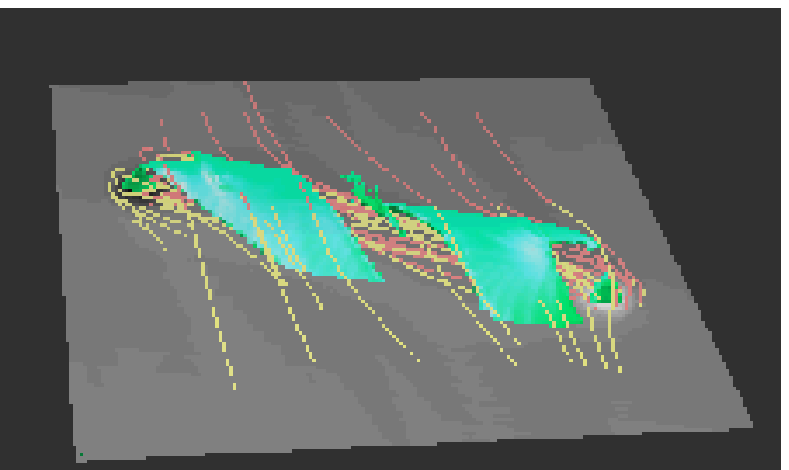}
\caption[]{\label{current_sheet.fig} Left, the situation where separator reconnection dominates
the dynamical evolution. Right, the situation at a later time where the field has started to disconnect
through separatrix-surface reconnection. The green isosurfaces indicate the locations of strong current,
while the field lines indicate the structure of the magnetic field.

}
\end{figure}

Experiments have been conducted with a number of different orientations of the overlying
magnetic field. The implications of this are, first, the distance the sources have
to be advected before the two flux domains start to interact vary with the overlying field angle; and second, 
so does the velocity perpendicular to the field in the flux domains. At all orientations of the overlying field, 
the same general evolution takes place.
As the two flux domain are forces into each other a separator current sheet is formed connecting the
two null points located on the bottom boundary. As the current magnitude reaches the numerical limit
magnetic {\it separator} reconnection starts and new field lines are created connecting the two sources. 
This process continues
until the two sources are well passed their point of closest approach and the two sources are well connected. 
Eventually current concentrations
start forming over the separatrix surface that encloses the connected flux resulting in {\it separatrix-surface}
reconnection that
disconnects this field. The two evolutionary phases are shown in \Fig{current_sheet.fig},
where the large isosurfaces represent the locations of the current sheets responsible for
the reconnection and the field lines indicate the local and global connectivity of the magnetic field.

\section{Magnetic Reconnection Rates}
An important factor of magnetic reconnection in three-dimensions relates to the speed at which the process
can take place. From two-dimensional magnetic reconnection (see 
Priest \& Forbes 2000 and references
there in) it is know that the speed of magnetic reconnection depends critically on the
boundary conditions of the investigated domain. Furthermore, most of the previous works have concentrated on
 steady-state situations. Here, the situation is different, the system is driven and it reacts to
the stresses created by the boundary motions and therefore finds it's own ``boundary conditions'' in the
language of 
Priest \& Forbes (1986) and in this sense finds its own steady state.

\begin{figure}
\center
\includegraphics[width=0.32\linewidth]{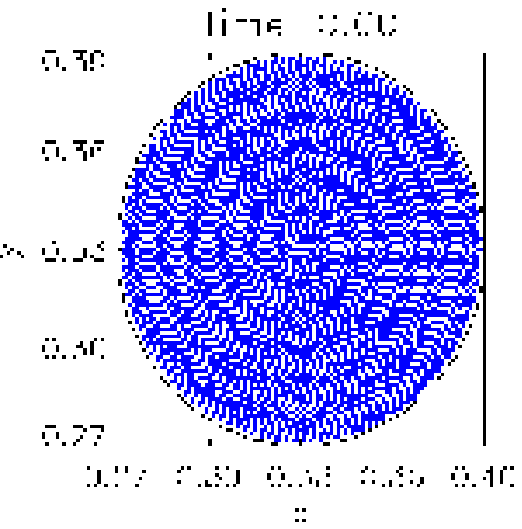}
\includegraphics[width=0.32\linewidth]{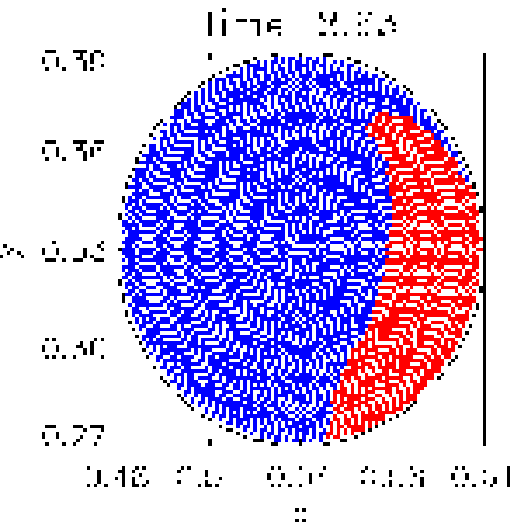}
\includegraphics[width=0.32\linewidth]{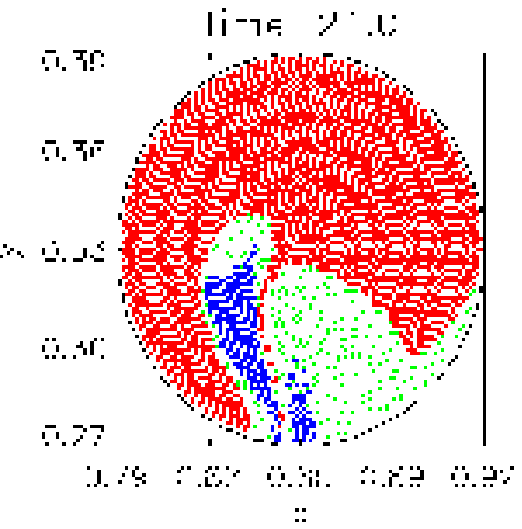}
\caption[]{\label{connectivity.fig} The field line connectivity at three different times. The
colour coding represents different types of field line connectivity: blue indicates initially
unconnected flux, red represents connected flux and green represents the disconnected flux.

}
\end{figure}

To investigate the rate of change in connectivity a large number of field lines have been traced from
the foot points of each source, \cite{Parnell_Galsgaard04}.
As the source structure is maintained in time,
each field line can simply be associated with a fixed amount of flux. Therefore as the field
lines change connectivity the amount of flux involved in each process
can be counted.  In \Fig{connectivity.fig}, three snapshots showing the
structure of the field line connectivity are shown. In these snapshots, blue represents the initially
unconnected flux, red indicates the field lines that connect the two sources and the green area is 
associated with field lines that have been disconnected.
These graphs show that the change in connectivity takes place in a non-trivial manner, providing a
complex mapping of the field line connectivity. In the potential extrapolation of the coronal
magnetic field by 
Close et al (2004a), it was shown that the field line mapping of individual
sources can connect to several neighbouring sources at the same time. From the mapping diagrams
shown in \Fig{connectivity.fig} it is seen that, even for this simple case, the
connectivity map contains three different types of field line connectivity.
This suggests that the mappings of Close at al (2004a) provide only a lower limit in
terms of complexity and that the added effects of the dynamical evolution may increase the
complexity of the field line connectivity of the coronal field even further.

\begin{figure}
\center
\includegraphics[width=0.47\linewidth]{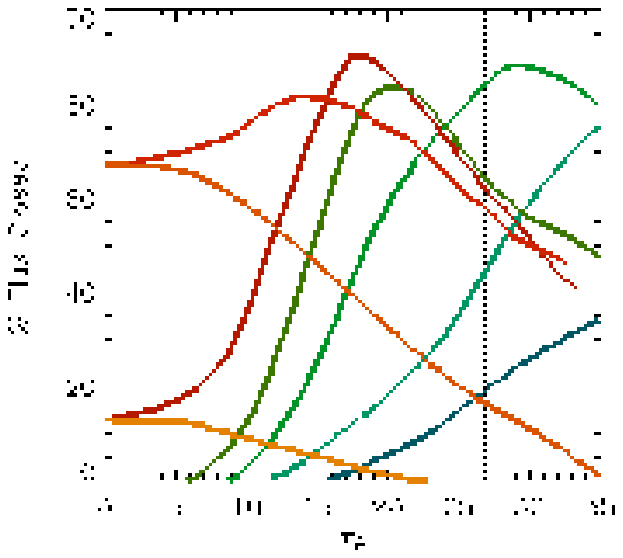}
\includegraphics[width=0.47\linewidth]{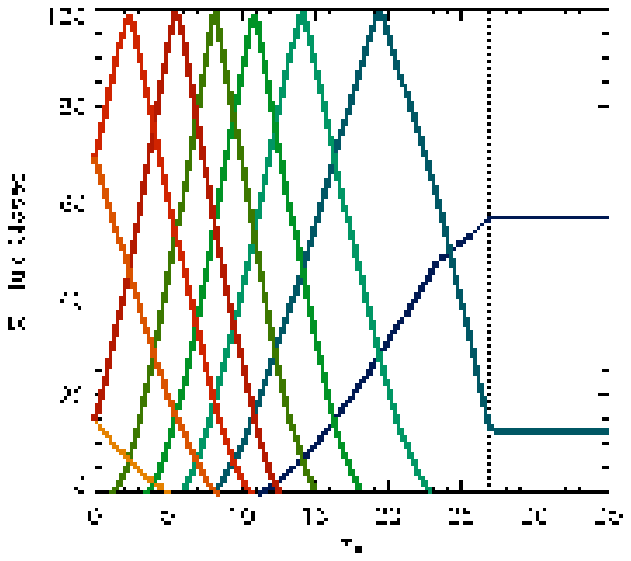}
\includegraphics[width=0.90\linewidth]{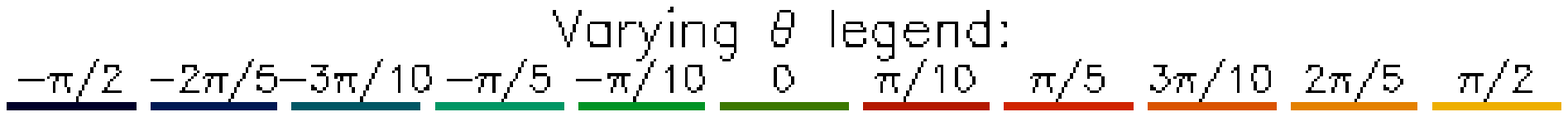}
\caption[]{\label{flux_change.fig}
Left, the change in connected flux as a function of time for the experiments
with different angles of the overlying horizontal magnetic field.
Right, the comparable graphs for the potential evolution of the experiments.
}
\end{figure}

Another important effect of the dynamical evolution is the fact that multiple reconnection
events take place at different spatial locations. This is clearly seen from the graphs showing
the change in connectivity as a function of time where it is found that the reconnection processes
connecting and disconnecting the flux systems, for a limited time interval, take place simultaneous.
In \Fig{flux_change.fig}
an example of the time depended evolution of the connected flux is shown. The left panel
shows the evolution in the situation where the magnetic field is assumed to always relax down to a potential
magnetic field configuration. In the right panel, the comparable evolution is shown for the
dynamical MHD development of the system.
Since both connecting and disconnecting take place at the same time, 
to derive a simple reconnection rate for the system both processes must be taken into account. 
The process of tracking the field line connectivity provides a
possibility for doing exactly this. The amount of flux belonging to any of the three classes
are therefore easily obtained as a function of time. Taking the time derivative of these and
normalising the result with regard to the characteristic Alfv{\'e}n velocity of the domain and
the total flux in a source, gives a dimensionless measure of the reconnection speed.
\begin{figure}
\center
\includegraphics[width=0.95\linewidth]{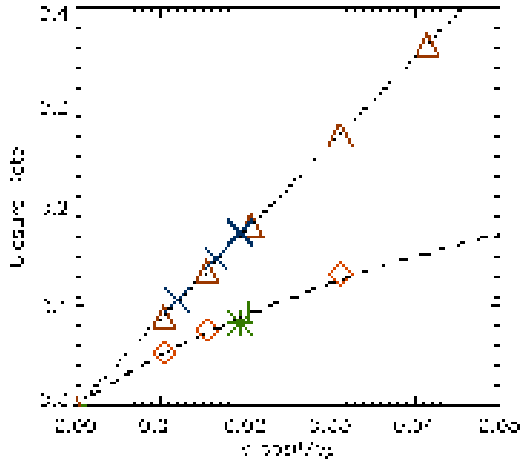}
\caption[]{\label{rec_rate.fig} The peak total reconnection rate for the potential evolution
(dotted line) and the dynamical evolution (dashed line). The symbols represent
the various individual experiments while the lines are lines of best fit given by \Eq{rec_rate.eq} and 
\Eq{pot_rec_rate.eq}, respectively.}
\end{figure}

From \Fig{flux_change.fig} it is seen that the fraction of connected flux between the two
sources depends both on the angle of the overlying magnetic field and the driving speed
and clearly on whether the evolution is potential or the dynamical MHD solution. 
The time derivative of these curves gives a measure of the rate of change of the connectivity,
which is equivalent to measuring a connecting reconnection rate. Similar graphs can be obtained for
the disconnecting process and hence the disconnecting reconnection rate. 
By comparing these phases it is found that the connecting and disconnecting
of the field in the potential case occurs at the same rate.
In the dynamical experiment, however, these processes are found to evolve at two different rates; 
with the peak connecting rate found to be
about twice that of the peak disconnecting rate. This difference in reconnection rates is a
consequence of the apparent stress of the magnetic field for the two process. In the connecting process
the two independent flux systems are forced into each other as the system responds to the
advection of the sources and the only way to release this stress is to generate a local
current sheet through which magnetic reconnection can take place.
In the following disconnecting process, reconnection takes place
between open overlying magnetic field and the connected field which is advected under this. 
Here, the stresses between the two flux regions are
less, as the overlying open field can simply be lifted up over the connected (dipole) field.
The developing separatrix current sheets are consequences of the increasing tangential change in
the field line orientation between these two flux systems and the pushing of them together by the 
magnetic tension and magnetic pressure forces in the two systems.

Adding the rates of connecting and disconnecting
gives the total reconnection rate at any one time. The result of this is
seen in \Fig{rec_rate.fig}, which shows both the potential (perfect, instantaneous) reconnection rate and the
comparable result for the dynamical evolution for all the conducted experiments. This graph doesn't
provide detailed insight to the reconnection process, but shows how that the accumulated peak
reconnection rate for the experiments depend, in a systematic way, on
the driving velocity perpendicular to the direction of the overlying magnetic field.
In \Fig{rec_rate.fig}, the dotted line represents the potential evolution of
the magnetic field, while the dashed line represents the dynamical evolution. The important
information contained in this diagram can be condensed to the following expressions for the
two reconnection rates:
\EQA
\label{pot_rec_rate.eq}
R_{pot} & = & 8.8 v_d \cos{\theta}, \\
\label{rec_rate.eq}
R_{dyn} & = & {{0.6 v_d \cos{\theta} }\over {v_d \cos{\theta}+0.11}},
\ENA
where $v_d \cos{\theta}$ is the driving velocity perpendicular to the direction of the overlying
magnetic field, which also represents the direction perpendicular to the developing current sheet.
This shows that for slow driving velocities the dynamic peak reconnection rate approaches the potential
rate, while for increasing driving velocities it approaches
a constant rate whereas the potential rate continues to grow linearly. 
For realistic driving velocities in the solar context, this indicates that the magnetic
reconnection rate, here, is fast, at just a factor of two less than the potential rate.

\section{Energy Considerations}

Taking the result from the previous section, namely that the reconnection is fast and within a factor
of two of that the required for the field to maintain a potential state,
one may assume that the energetics of the system could simply be represented using a potential
model \cite{Galsgaard_Parnell04}.

\begin{figure}
\center
\includegraphics[width=0.47\linewidth]{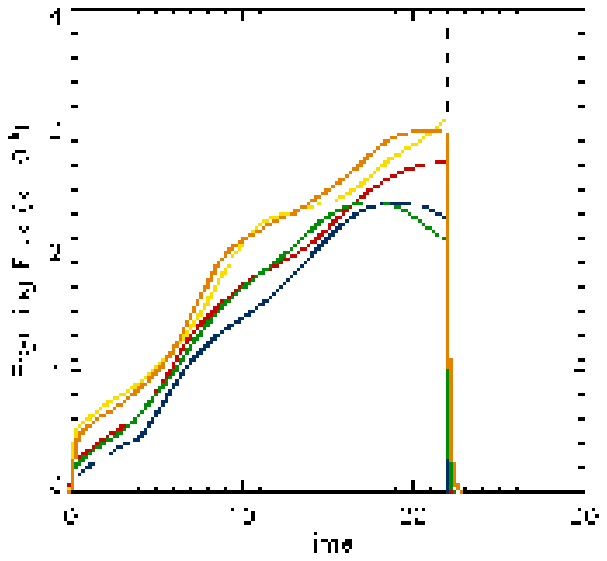}
\includegraphics[width=0.47\linewidth]{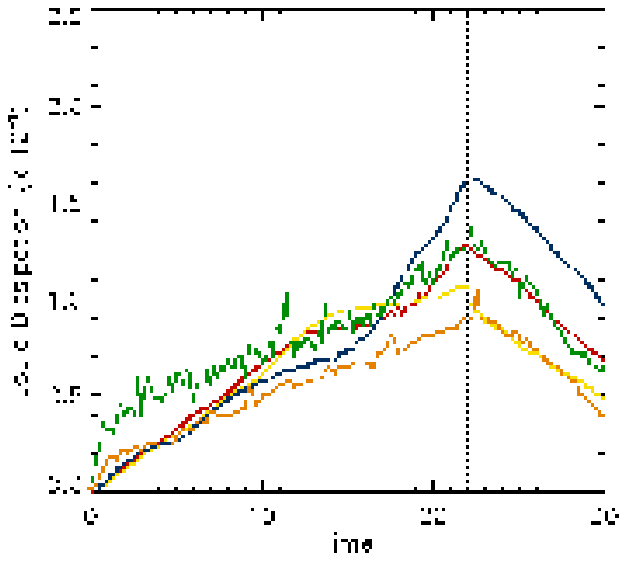}
\includegraphics[width=0.85\linewidth]{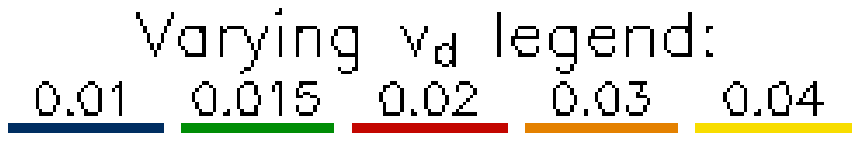}
\caption[]{\label{pf.fig} Left, the pointing flux for five experiments with different driving
velocity. The graphs are scaled according to \Eq{pf.eq} to give the same absolute value
as a function of advection distance. Right, the average Joule dissipation.
As a first approximation, the energy released from all
the experiments, over a given advection distance of the sources, may be considered the roughly same.
The colours of the lines represent different driving velocities.}
\end{figure}

If the driving velocity is slow and has a duration that is long compared to the Alfv{\'e}n
crossing time then the Poynting flux input can be estimated by,
\EQ
\label{pf.eq}
P_{f} \approx \mu {{B_n v_d^2 t_d}\over L},
\EN
where $B_n$ is the normal component of the magnetic field, $v_d$ is the driving velocity, $t_d$ is the
duration of the driving and $L$ is the distance to the anchored (fixed) end
of the field line.
This reveals that the energy input scales quadratically with driving speed and linearly with
duration. Integrating \Eq{pf.eq} over a fixed advection distance suggests that the
same amount of energy should be injected to the magnetic configuration independent of the driving velocity. 
These results have been tested by conducting a series of experiments with different driving velocities. 
Scaling the amplitudes of the Poynting flux to the
advection distance indicates that the results follow the simple scalings obtainable from \Eq{pf.eq} to within
a velocity dependent initial offset, left panel of \Fig{pf.fig}.
The initial difference is due to the change in ratio between the driving velocity and the local
Alfv{\'e}n velocity and holds as long as the information can propagate freely away from the driving
boundary. It is only when the initial perturbation of the field lines, which is generated when the driving 
is first switched on, reaches another anchor point of the field line that the effect of the angle of the field 
(caused by the driving) start to be felt leading to a linear increase in time of the Poynting flux according 
to \Eq{pf.eq}.

If the same amount of energy is injected into the domain for a given advection distance regardless of 
driving speed, then in each case the same amount of energy is available for heating the plasma through dissipation.
The dynamical evolution of the various cases follow similar patterns and so it is not unreasonable to 
expect the energy release to be of comparable magnitude.
The Joule dissipation for the numerical domain is plotted as a function of advection distance and
scaled to the same driving time in right panel of \Fig{pf.fig}. 
This panel shows that only minor differences in the energy release exist between the 
different cases over the same advection distance, and that most
of these can be accounted for by the differences in the ratio between the Alfv{\'e}n crossing
time and the driving time for a given advection distance. The effect of the
energy release, on the other hand, depends linearly on the timescale over which it is
released. The shorter the timescale the greater the impact of the energy release.

The amount of Poynting flux and Joule dissipation cannot be estimated using potential models, as
they depend critically on the magnetic fields reaction to the imposed driving, and through this on the
change in angle of the local magnetic field lines stressed by the boundary
driving. This angle is a result of a balanced between the geometry of the magnetic field, the local current
concentrations and the efficiency of the magnetic resistivity.

\section{Plasma Structure Of The Energy Release}

\begin{figure}
\center
\includegraphics[width=0.47\linewidth]{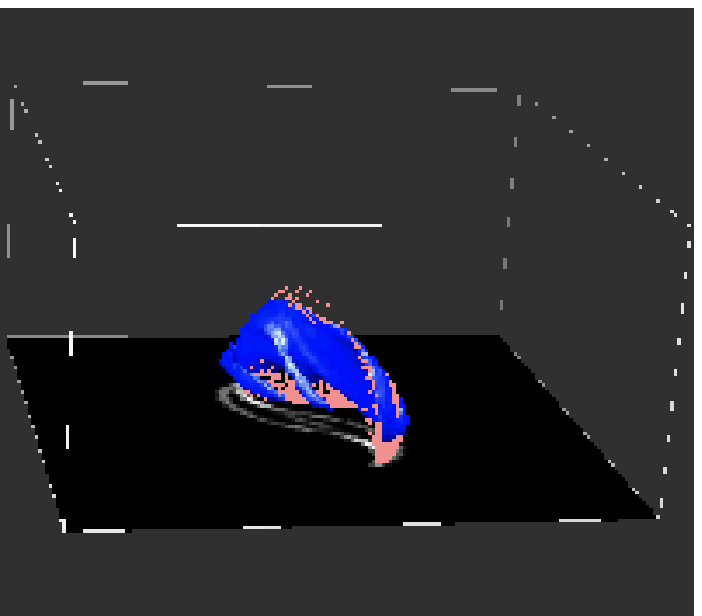}
\includegraphics[width=0.47\linewidth]{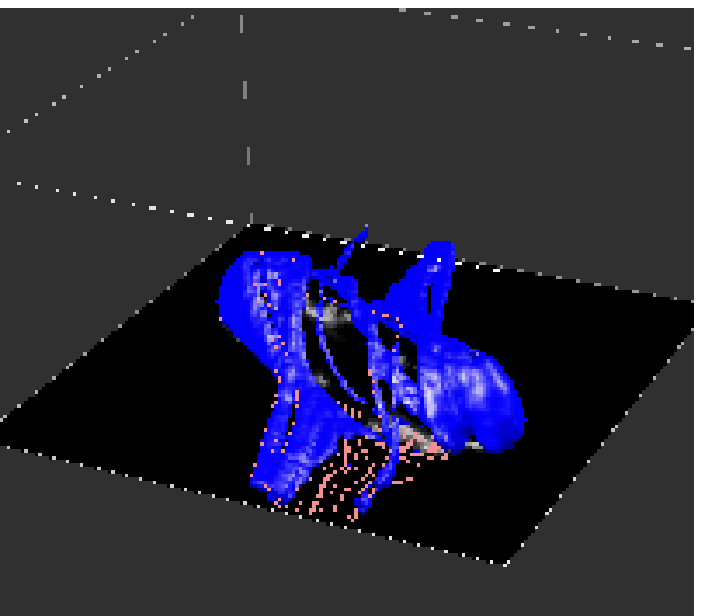}
\includegraphics[width=0.47\linewidth]{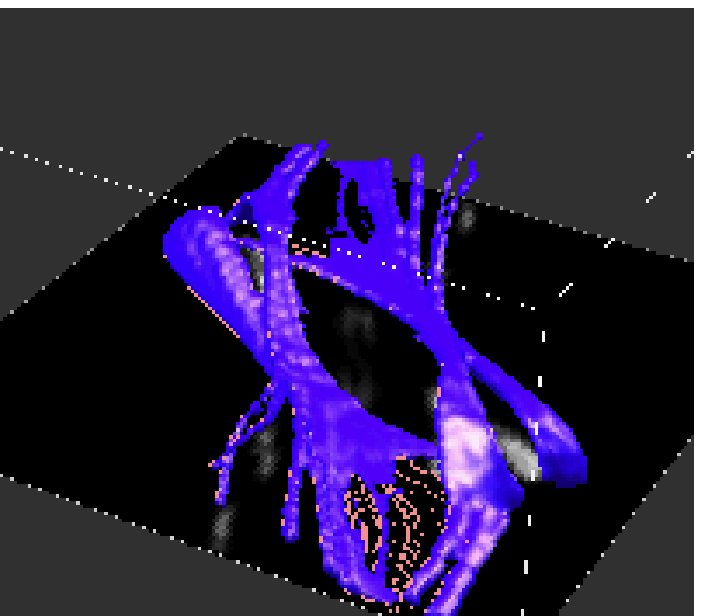}
\includegraphics[width=0.47\linewidth]{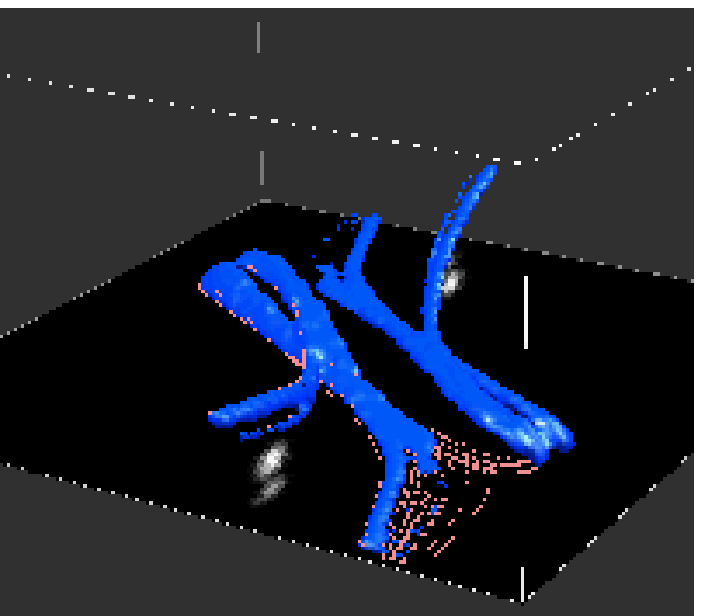}
\caption[]{\label{heating.fig}The four panels show the structure of the heated plasma at
different times during the evolution of the flux systems in one experiment. Only the regions with temperatures
above the initial coronal temperature are shown.}
\end{figure}

The models used in these investigations have been setup to investigate the effects of magnetic
reconnection in a simple magnetic structure. But they can also be used to attempt to
investigate the change in the visual appearance of heated magnetic structures as the type of reconnection
changes with time.
The best and only realistic way to do this is to include: i) a more detailed model atmosphere;
ii) the effects of optically thin radiation and anisotropic heat conduction in the energy equation.
The disadvantage with this is the requirement for background heating to balance the radiation and conduction 
in the initial configuration which one cannot easily justify in
the given system. Instead, a more simplistic approach to illustrate the effect has been
adopted here. From the field line tracing mentioned above it is straight forward to identify the starting points
of the field lines that change connectivity between the snapshots.
A realistic assumption to then make is that the energy released in the reconnection event scales with the magnetic
flux represented by the field line. Distributing this energy evenly along the field lines,
a local temperature increase of the plasma may then be estimated with the inclusion of a vertical density 
profile along the field lines. Here, the plasma density is assumed to decreases rapidly with height following
a simple hyperbolic tan function. This simple model emulates a photosphere at the driving boundary
follow by a shallow transition region, while most of the domain contains a coronal plasma density.
Then by following the reconnected field lines, local temperature increases can be
calculated which, when all contributions are summed, reveal a few interesting
conclusions as the system evolves in time (\Fig{heating.fig}).
\begin{itemize}
\item The initial connecting of the magnetic field between the two sources gives rise to a
compact structure outlined by field lines that link the two sources.
\begin{itemize}
  \item Note, that a limitation here is the problem with identifying the newly reconnected open field lines which 
   are not identified. These coronal field lines may, however, be so long that the energy distributed along 
   them from the reconnection has only limited impact on rising their temperature.
\end{itemize}
\item When the field starts to disconnect the field lines extend from the sources to the open coronal field again
   and the energy is distributed along field lines that might reach far away from the reconnection site.
\item Furthermore, the disconnecting process is not equally active all across the whole separatrix surface, 
   but is more stochastic in nature leading to several fingers of field lines leading away from the sources.
\end{itemize}
Taking the simplicity of the approach into account it is clear that to fully see the thermal,
and therefore observational appearance of the reconnecting structure, the problem has to be solved
consistently including the effects of optical thin radiation and anisotropic heat conduction.
An interesting, but difficult task to handle consistently and clearly a problem for future investigations.

\section{Summary}

The numerical experiments discussed in this paper investigate a simple situation where two
initially unconnected flux sources are forces into each other by an imposed velocity profile
on the photospheric boundary. From the experiments the following main conclusions
arises:

\begin{itemize}
\item The reconnection initially leading to the connection of the two sources occurs in the form of separator
reconnection along the degenerate separator line connecting two magnetic nulls located in the photospheric plane.

\item The disconnecting of this flux takes place through separatrix-surface reconnection.

\item Mapping the changes in connectivity of one source, as a function of time, shows a complicated
connectivity map with a clear overlap between initially unconnected, connected and disconnected flux.

\item Reconnection speeds for the two process have been obtained showing that separator reconnection
connecting the field is about twice as fast as the separatrix-surface reconnection that disconnects the field.

\item The dynamical model is found to be about a factor of two slower, for realistically slow driving velocities, 
than a potential model in which instantaneous perfect reconnection occurs.
In other words, {\sl the reconnection is fast}!

\item The Poynting flux injected through the imposed driving depends mainly on the advection distance
and only to a small degree on the driving velocity. This result cannot be obtained using potential
models.

\item Similarly, the time integrated Joule dissipation between the various experiments is comparable for a given
advection distance. This implies that the effect of the energy deposition depends critically on the
timescale of the energy release.

\item The reconnected magnetic field lines form a compact structure linking
the two flux sources, giving rise to a very local area of plasma above the background
coronal temperature.

\item In the disconnecting phase the reconnected (heated) field lines create a complicated web of hot 
filaments extending into the corona.

\end{itemize}

In terms of observational appearance of the flux interaction the applied model is very simple and
more detailed investigations, including a more realistic treatment of the energy equation are
required. This is work that has to be done, if we want to conduct detailed comparisons
between numerical models and observations.

\section*{Acknowledgments}
KG would like to thank PPARC and Carlsberg for financial support over various
periods of this work in the from of two
fellowships. CEP would like to thank PPARC for support as an Advanced Fellow. The computational analysis for this
paper was carried out on the joint SRIF and PPARC funded linux cluster, copson, in St. Andrews.

\bibliographystyle{aa}
\bibliography{esapub}

\end{document}